\documentclass[reprint,prb,superscriptaddress,amsmath,amssymb,aps
]{revtex4-1}

\usepackage{bm}%
\usepackage[colorlinks=true,linkcolor=blue]{hyperref}%
\expandafter\ifx\csname package@font\endcsname\relax\else
 \expandafter\expandafter
 \expandafter\usepackage
 \expandafter\expandafter
 \expandafter{\csname package@font\endcsname}%
\fi
\usepackage{graphicx}
\usepackage{dcolumn}

\begin{document}

\title{A L\'evy flight for electrons in graphene: superdiffusive-to-diffusive transport transition}

\author{Diego B. Fonseca}
\affiliation{Departamento de F\'{\i}sica, Universidade Federal Rural de Pernambuco, Recife - PE, 52171-900, Brazil}

\author{Luiz Felipe C. Pereira}
\email{luiz.cpereira@ufpe.br}
\affiliation{Departamento de F\'{\i}sica, Centro de Ciências Exatas e da Natureza, Universidade Federal de Pernambuco, Recife - PE, 50670-901, Brazil}

\author{Anderson L. R. Barbosa}
\email{anderson.barbosa@ufrpe.br}
\affiliation{Departamento de F\'{\i}sica, Universidade Federal Rural de Pernambuco, Recife - PE, 52171-900, Brazil}

\begin{abstract}
In this work we propose an electronic L\'evy 
glass, analogous to a recent optical realization. 
To that end, we investigate the transmission of electrons in graphene nanoribbons in the presence of circular electrostatic clusters, whose diameter follow a power-law distribution. 
We analyze the effect of the electrostatic clusters on the electronic transport regime of the nanoribbons, in terms of its diffusion behavior. 
Our numerical calculations show that the presence of circular electrostatic clusters induces a transition from L\'evy (superdiffusive) to diffusive transport as the energy increases. 
Furthermore, we argue that in our electronic L\'evy 
glass, superdiffusive transport is an exclusive feature of the low-energy quantum regime, while diffusive transport is a feature of the semiclassical regime.
We thus attribute the observed transition to the chiral symmetry breaking, once the energy moves away from the Dirac point of graphene.
\end{abstract}


\maketitle

\section{Introduction} 

Graphene is a unique platform to emulate wave optics by electronic phenomena, since its linear dispersion relation at low excitation energy coincides qualitatively with the photon's dispersion [\onlinecite{Caridad2016,PhysRevLett.106.176802,PhysRevLett.120.124101,doi:10.1126/science.1138020}]. 
This linear dispersion is due to the honeycomb lattice of graphene, which can be seen as a triangular Bravais lattice with a two-atom basis [\onlinecite{Novoselov2004,Novoselov2005,RevModPhys.81.109}]. 
In graphene, charge carriers behave as massless relativistic Dirac fermions, and the lattice preserves chiral symmetry. 
These unique electronic properties give rise to  Klein tunneling [\onlinecite{Katsnelson,PhysRevE.106.054127}], where  massless fermions can tunnel through a potential barrier with null reflection probability.

In this context, Ref. [\onlinecite{PhysRevB.87.155409}] reported the electronic analogue of Mie scattering in a graphene superlattice imbibed in a cylindrical electrostatic potential [\onlinecite{PhysRevLett.102.226803,Gutierrez2016}]. 
Mie scattering is an optical phenomenon which takes place when light waves are elastically scattered by spherical or cylindrical objects.
Furthermore, Ref. [\onlinecite{https://doi.org/10.1002/pssb.201552119}] studied the effects of a regular array of electrostatic quantum dot clusters (EQDC) in otherwise pristine graphene nanoribbons, which induces a local deformation of the on-site potentials.
Both theoretical predictions were confirmed in Ref. [\onlinecite{Caridad2016}], which presented an experimental demonstration of an electronic analogue of Mie scattering by using a graphene superlattice as a conductor imbibed into a regular EQDC array. 
They were followed by other relevant work on the effect of circular electrostatic potentials in graphene [\onlinecite{sadra,Walls2015,Walls2016,PhysRevB.105.165404,Li2021,Sadrara_2022}].

The experimental setup of Ref. [\onlinecite{Caridad2016}], inventively, reminds one of a remarkable optical wave transport work: {\it A L\'evy flight for light} [\onlinecite{bart}].
L\'evy flights are a particular class of non-Gaussian random walks in which a heavy-tailed (power-law) distribution describes the step length during the walk [\onlinecite{zaburdaev2015levy}].
Those flights are present in different fields of science such as the migration pattern of animals [\onlinecite{10.1371/journal.pcbi.1005774,PhysRevLett.124.080601}], transport in turbulent flows [\onlinecite{PhysRevLett.71.3975}], optical wave transport [\onlinecite{bart,Mercadier2009,PhysRevLett.113.233901,PhysRevA.85.035803,PhysRevResearch.3.023035,Li:19}] and electronic transport [\onlinecite{PhysRevLett.81.1274,PhysRevLett.117.046603,PhysRevLett.123.195302,PhysRevE.99.032118,PhysRevE.106.054127}].
L\'evy flights lead to superdiffusive transport, which is characterized by a mean-square displacement growing faster than linear with time, i.e., $\langle x^2 \rangle \propto t^{\gamma_D}$, where $\gamma_D > 1$.
Meanwhile, for $\gamma_D=1$ we recover the regular diffusive transport regime  [\onlinecite{zaburdaev2015levy,PhysRevX.7.021002,https://doi.org/10.48550/arxiv.2106.13892}]. 

Ref. [\onlinecite{bart}] developed an optical material in which the transport of light is governed by L\'evy statistics: a L\'evy glass.
The material was fabricated by suspending titanium dioxide microspheres in sodium silicate, where the diameter of the microspheres followed a heavy-tailed distribution. 
The suspended microspheres are a scattering material with fractal structures, where the larger particles give origin to Mie scattering. \textcolor{black}{Furthermore, since the configuration of microspheres is fixed in the Lévy glass, subsequent steps are correlated, in contrast to a standard  Lévy walk, where the steps are all uncorrelated [\onlinecite{PhysRevE.85.021138}].} 
The authors of Ref. [\onlinecite{bart}] observed that when the light is transmitted through the L\'evy glass, it shows superdiffusive transport instead of regular diffusive transport.
Therefore, the average transmission coefficient as a function of the device length $L$ follows [\onlinecite{barthelemy2009anomalous,bart,PhysRevE.85.021138}]
\begin{equation}\label{eq:alpha}
    \langle T \rangle = \frac{1}{1 + \left(L/\ell\right)^{\gamma}},
\end{equation}
where $\ell$ is the mean free path. 
When $\gamma = 1 $, we have the usual behavior giving rise to regular diffusive transport. 
Whereas, when $\gamma < 1 $, we have a slow decay of the transmission characterizing a superdiffusive transport regime. 

\begin{figure}
    \centering
    \includegraphics[clip, width=1.0\linewidth]{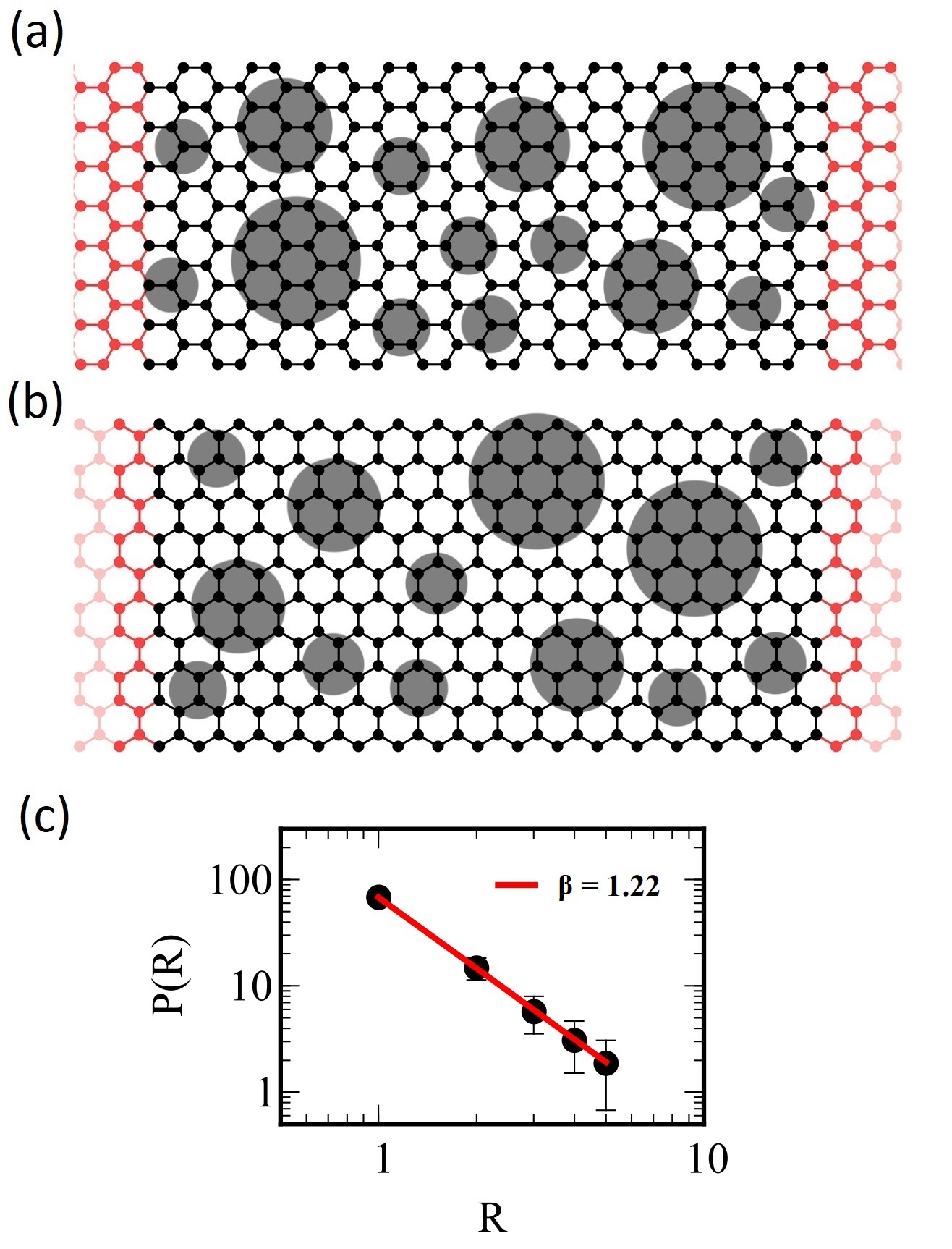}
    \caption{Illustration of (a) AGNR and (b) ZGNR connected to two leads (red). Grey circles represent EQDC. (c) Histogram of cluster radii (symbols); the solid line is a fit obtained from Eq. (\ref{PR}) with  $\beta = 1.22\pm0.01$.}
    \label{fig:rede_1}
\end{figure}

In order to emulate the optical L\'evy flight [\onlinecite{bart}] with electrons, we propose an electronic L\'evy glass (ELG) and develop realistic numerical calculations of the electronic transport through graphene nanoribbons imbibed into the EQDC, whose radii follow a heavy-tailed distribution, as shown in Fig. \ref{fig:rede_1}. 
We find that our ELG show a L\'evy-to-diffusive transport transition as a function of the Fermi energy, which is a particular feature of the graphene honeycomb lattice. 
Furthermore, our analysis indicates that the most probable cause of the observed transition is the breaking of chiral symmetry, once the energy moves away from the Dirac point of graphene.
Finally, we believe that the L\'evy-to-diffusive  transition can be experimentally verified by an adaptation of the experimental setup in Ref. [\onlinecite{Caridad2016}].

\section{ Microscopic Model}

We study the electronic transmission through graphene nanoribbons connected to two leads, and imbibed into randomly located EQDC, as illustrated in  Fig. \ref{fig:rede_1}.
The scattering matrix describing the electronic transport through the nanoribbons is given by  [\onlinecite{datta}]
\begin{equation}
    S = 
    \begin{bmatrix}
       {r} &{t'}\\
        {t} &{r'}
    \end{bmatrix},
\end{equation}
where ${t}({t'})$ and ${r}({r'})$ are the transmission and reflection matrix blocks, respectively. 
The transmission coefficient can be calculated from the Landauer-B\"uttiker relation
\begin{equation}
    T = {Tr}[{t}{{t}}^\dagger], \label{T}
\end{equation}
which is valid in the linear response regime. 
Numerical calculations of the transmission coefficient were performed with KWANT [\onlinecite{kwant}], which is a Green’s function–based algorithm within the tight-binding approach. 

The tight-binding Hamiltonian for graphene is given by
\begin{equation}
    \hat{H} = -t_0\sum_{\left \langle i,j \right \rangle}  c_{i}^\dagger c_{j} +  \sum_{i} \epsilon_i c_{i}^\dagger c_{i},\label{H}
\end{equation}
where the indices $i$ and $j$ run over all lattice sites and $\left \langle i,j \right \rangle$ denotes first nearest neighbors. The first term in $\hat{H}$ represents
the usual electron hopping between lattice sites, $c_{i}$ $(c_{i}^{\dagger})$ are the annihilation (creation) operators and $t_0$ is the hopping energy, which has a typical value of 2.7 eV [\onlinecite{RevModPhys.81.109}].
The second term in the Hamiltonian is the electrostatic potential induced by the EQDC. 
Therefore, the on-site electrostatic potential $\epsilon_i$ will be $\epsilon_i = V$ when the site is inside the quantum dot area, and $\epsilon_i = 0$ otherwise.

{The Hamiltonian in Eq. (\ref{H}) is identical to the one in Ref. [\onlinecite{https://doi.org/10.1002/pssb.201552119}], which studied the effects of a regular array of EQDC in otherwise pristine graphene nanoribbons, and motivated the experiment developed in Ref. [\onlinecite{Caridad2016}]. This is a deliberate choice which allows us to probe the effect of the Lévy-disorder on electronic transport, and to perform a direct comparison with a periodic disorder.
Furthermore, Ref. [\onlinecite{https://doi.org/10.1002/pssb.201552119}] also did not consider edge passivation in the nanoribbons since it should not influence electronic transport significantly in the presence of EQDC. 
Finally, the EQDC are not expected to modify the hopping parameter significantly, and thus we consider the same hopping energy inside and outside the clusters, also in line with Ref.  [\onlinecite{https://doi.org/10.1002/pssb.201552119}].}

Figs. \ref{fig:rede_1}(a) and (b) illustrate armchair (AGNR) and zigzag (ZGNR) graphene nanoribbons, respectively, imbibed into the EQDC.
In order to build the ELG, we follow four steps: 1) randomly select both a point on the lattice, which will be the center of the quantum dot, and the radius $R$ of the quantum dot; 2) assign all sites inside the quantum dot area a constant electrostatic potential value $\epsilon_i = V$; 3) randomly select a new lattice point and radius: if the new quantum dot overlaps with a pre-existing one, begin step 3 again, otherwise go to step 2; 4) stop after 5000 consecutive failed attempts to introduce a new quantum dot. 
The maximum radius of the quantum dot is limited to one eighth of the lattice width without any loss of generality. 
In fact, this restriction is common in studies involving L\'evy distributions because of the divergence of its second moment [\onlinecite{zaburdaev2015levy}].

A L\'evy distribution is characterized by the probability density of a random variable $P(R)$, which has a power-law tail [\onlinecite{zaburdaev2015levy,PhysRevE.96.062141,PhysRevE.99.032118}]. 
The probability density is given by
\begin{equation}
    P(R) \propto \frac{1}{R^{\beta+1}},\label{PR}
\end{equation}
where $0 < \beta < 2$. 
If $0 < \beta < 1$, the first and second moments of $P(R)$ diverge because of heavy-tails, while for $1 \le \beta < 2$ only the second moment diverges.
Fig. \ref{fig:rede_1}(c) shows the radii histogram obtained from $10^4$ different AGNR samples with width $W_A=49.5a_0$ and length $L_A=51.4a_0$, where $a_0 = 2.49 \text{ \AA}$ is the graphene lattice constant. 
The histogram can be adjusted to the power law given in Eq. (\ref{PR}) with an exponent $\beta = 1.22\pm0.01$. 
Because of the L\'evy distribution behavior, a set of quantum dots with radius $R$ occupies approximately the same area as another set of quantum dots with a different radius. 
Similar behavior was reported by Ref. [\onlinecite{bart}]. 
Furthermore, in our simulations the fraction of the nanoribbon area occupied by quantum dots is $42.24\%\pm0.03$,  and it remains unchanged for different device lengths $L$, widths and edge types.

\section{Results}

To understand the effect of the EQDC on the average transmission coefficient, we build an ensemble of ELG for AGNR and ZGNR.
We kept fixed the widths, $W_A=49.5a_0$ and $W_Z = 49.6a_0$, and the electrostatic potential $V=0.25t_0$, while varying the length up to $L_A=1050.7a_0$  and $L_Z=1050.5a_0$, respectively. 
Besides, for future comparison between honeycomb and square lattices, we also build a set of square lattices, which we refer to as two-dimensional electron gas (2DEG), with $W_{2D}=50a_0$ and $L_{2D}$ up to $1050a_0$.

\begin{figure*}[htbp!]
    \centering
    \includegraphics[width=0.65\linewidth]{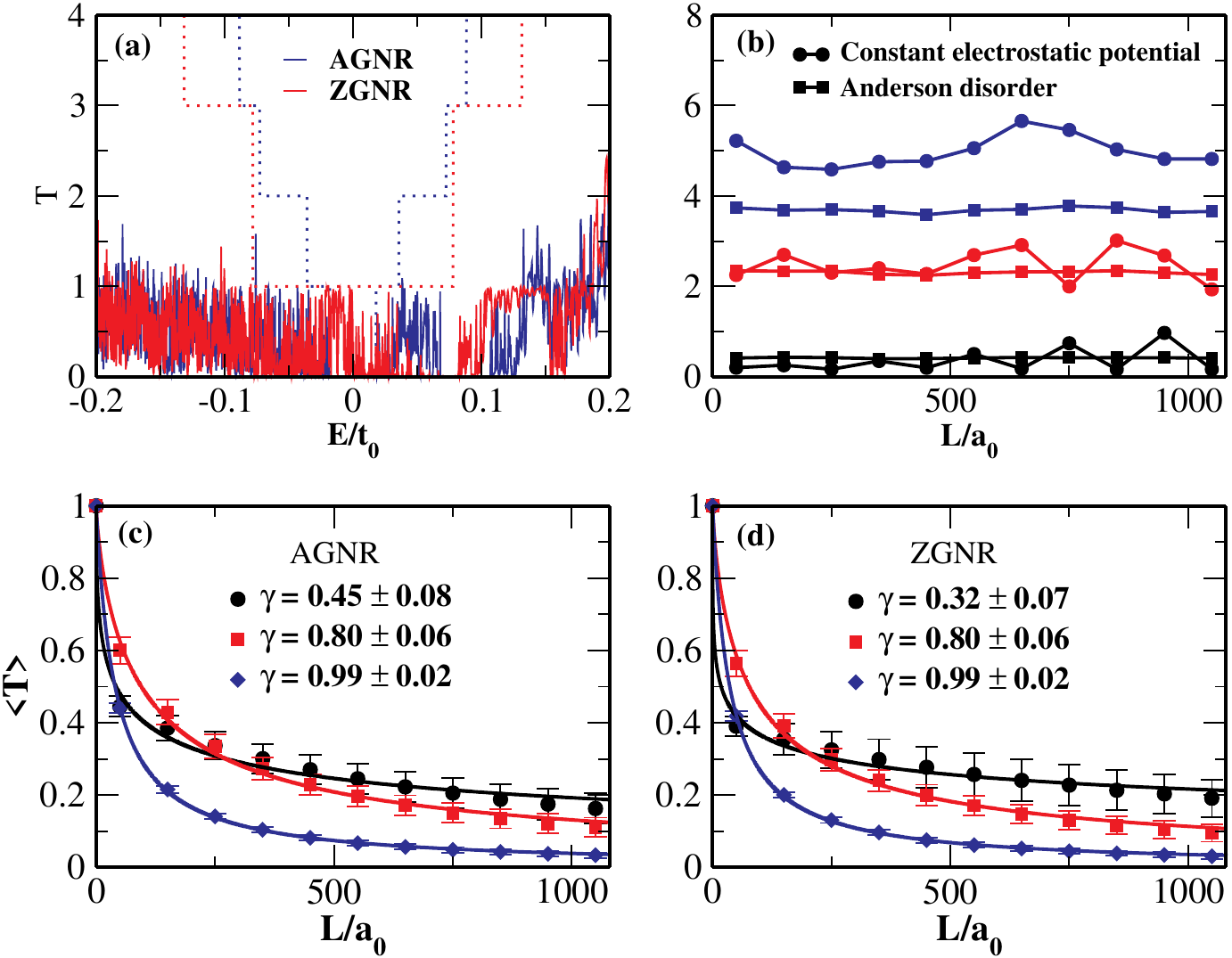}
    \caption {(a) Transmission through the ELG as a function of Fermi energy (solid lines); dotted lines are for the pristine nanoribbons. (b) Transmission as a function of length $L$ for three different energies (increasing from bottom to top). The circles are for AGNR with constant electrostatic potential $\epsilon_i=V$ in all sites, while squares are for AGNR with Anderson disorder. (c) Average transmission as a function of length $L$ for AGNR, with Fermi energy $E/t_0 = 0.2$ (circles), $0.34$ (squares) and $1.2$ (diamonds); (d) same for ZGNR, with $E/t_0 = 0.2$ (circles), $0.4$ (squares) and $1.2$ (diamonds). Solid lines in (c) and (d) are fits obtained from Eq. (\ref{eq:alpha}).}
    \label{fig:T_L}
\end{figure*}


{We begin by plotting the transmission coefficient as a function of Fermi energy in Fig. \ref{fig:T_L}(a) for $L_A=1050.7a_0$ and $L_Z=1050.5a_0$. 
For the ELG, the transmission (solid line) presents strong fluctuation due to EQDC, and is generally smaller when compared to the pristine nanoribbons (dotted lines).}  
Meanwhile, Fig. \ref{fig:T_L}(c) shows the average transmission through the AGNR as a function of length $L$ for three different Fermi energies, $E/t_0 = 0.2$, $0.34$ and $1.2$. 
The average transmissions (symbols) were calculated from a set of 6000 different ELG and  conveniently normalised to range between 0 and 1. Fig. \ref{fig:T_L}(d) is the equivalent for a ZGNR with $E/t_0 = 0.2$, $0.4$ and $1.2$. 

The average transmission of Figs. \ref{fig:T_L}(c) and (d) decrease as power-law functions of length $L$.
Therefore, we can fit the data with Eq. (\ref{eq:alpha}) and obtain the exponent $\gamma$ for each Fermi energy.
The exponent characterizes the diffusive ($\gamma=1$) and superdiffusive ($\gamma < 1$) transport regimes. 
Figs. \ref{fig:T_L}(c) and (d) show that for high Fermi energy ($E/t_0 = 1.2$) the exponent is $\gamma \simeq 1$ for both AGNR and ZGNR. 
Meanwhile, when we decrease the Fermi energy the exponent decreases to $\gamma\simeq 0.5$, which indicates a non-trivial L\'evy-to-diffusive transport transition as the Fermi energy increases.

Before proceeding with our investigation, we could ask ourselves: {\it Is the average transmission behavior shown in Figs. \ref{fig:T_L}(c) and (d) really due to the EQDC?} 
To answer this question, we first developed a numerical calculation of the transmission through nanoribbons with a constant electrostatic potential $\epsilon_i=V$ in all sites, at three different Fermi energies. 
Fig. \ref{fig:T_L}(b) shows that for the AGNR the transmission remains almost constant as function of length (circles).
As a second verification, we ran a numerical calculation with a typical Anderson disorder  [\onlinecite{nomura2007topological, lewenkopf2008numerical}]. 
In this case, the electrostatic potential $\epsilon_i$ varies randomly from site to site according to a uniform distribution in the interval $\left(-V/2,V/2\right)$. 
Again, we found that for the AGNR the transmission remains almost constant as a function of length, Fig. \ref{fig:T_L}(b) (squares).
The same behavior was observed for a ZGNR, which leads to the conclusion that the L\'evy-to-diffusive  transition is indeed a consequence of the EQDC, {and not due to some arbitrary electrostatic potential.}

\begin{figure*}[htbp!]
    \centering
    \includegraphics[width=0.65\linewidth]{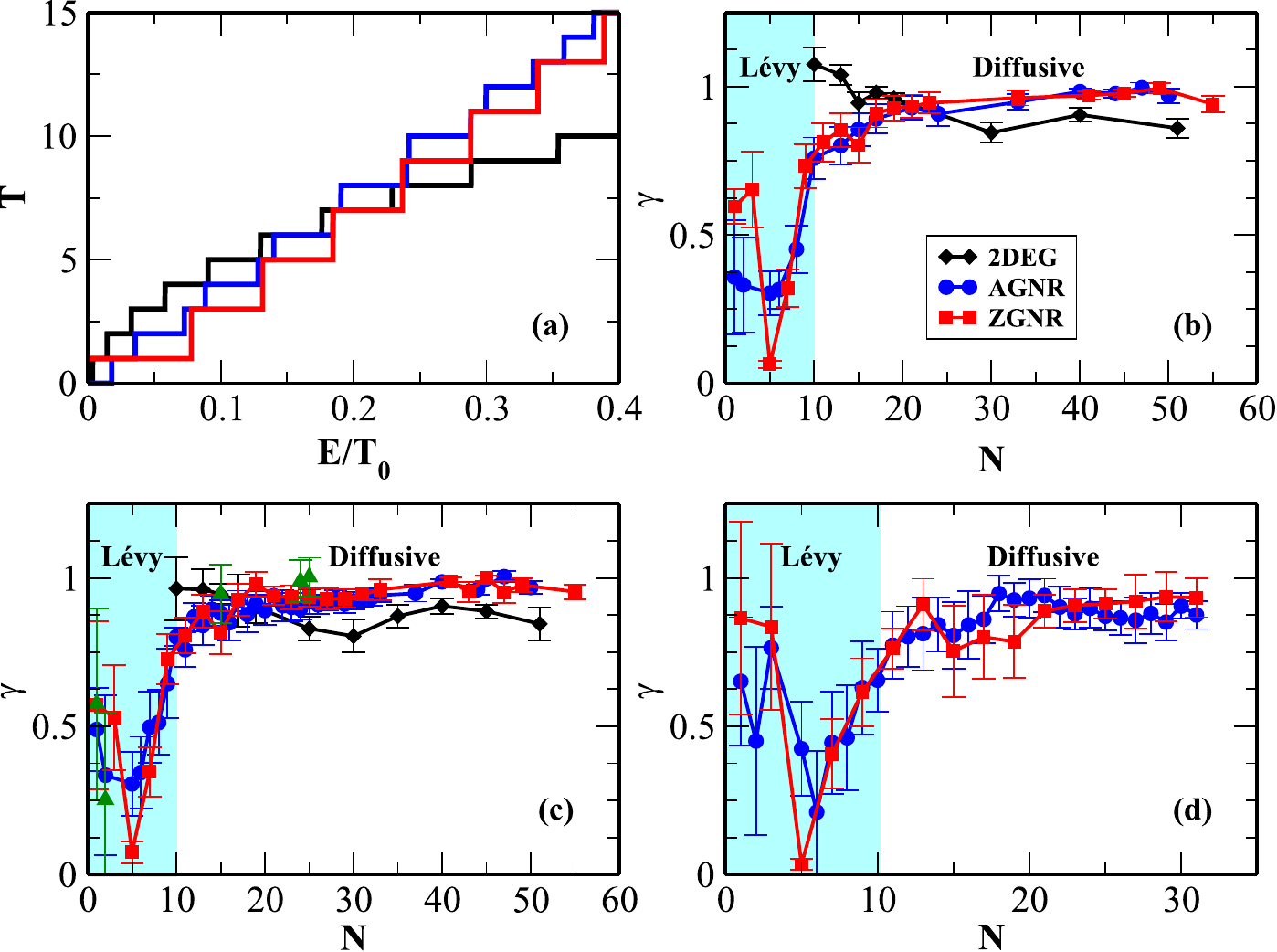}
    \caption{(a) {Transmission versus Fermi energy. The value at the plateaus corresponds to the number of channels $N$ in the AGNR, ZGNR and 2DEG leads}. The exponent $\gamma$ as a function of $N$ for: (b) a set of ELG samples {with $W_A=49.5a_0$, $W_Z = 49.6a_0$ and $W_{2D}=50a_0$}; (c) just one ELG sample {with $W_A=49.5a_0$, $W_Z = 49.6a_0$ and $W_{2D}=50a_0$. (Triangular symbols are for an AGNR device with second-nearest-neighbor hopping)}; (d) one ELG sample {with $W_A= 59.5a_0$ and $W_Z = 60.0a_0$}. Solid lines in (b), (c) and (d) are just to guide the eyes.}
    \label{fig:alpharules}
\end{figure*}

In order to characterize the L\'evy-to-diffusive  transition, it is convenient to relate the Fermi energy to the number of channels in the nanoribbon $N$. This dimensionless integer corresponds to the number of propagating wave modes in the ribbon, which is proportional to both the width $W$ and the Fermi vector $k_F$ through $N = k_FW/\pi$. {Since our aim is to study electronic localization, we keep $W$ fixed and increase the graphene length $L$ tuning $N$ only by the Fermi energy. If the width $W$ is kept fixed, the electronic structure of each nanoribbon remains unchanged during the numerical calculation. Nonetheless, we have verified that the transition does not depend on $W$, as discussed below.} 

Fig. \ref{fig:alpharules}(a) shows the transmission coefficient of pristine AGNR, ZGNR and 2DEG as a function of the Fermi energy.
The value at the plateaus corresponds to the number of channels, i.e. $T(E)=N$. 
For the AGNR (blue line), we have a relation between $E/t_0 = 0.02, 0.05, 0.08, \cdots$ and $N = 1, 2, 3, \cdots$, similarly for ZGNR (red line) and 2DEG (black line).
Knowing the relation between the Fermi energy and the number of channels, we can study how the exponent $\gamma$ varies as a function of $N$.

Fig. \ref{fig:alpharules}(b) shows the exponent $\gamma$ as a function of $N$ ranging from 1 to 55 for the AGNR, ZGNR and 2DEG. 
When $N > 10$,  Fig. \ref{fig:alpharules}(b) presents a plateau close to $\gamma=1$ for the three types of ELG, which indicates  regular diffusive  transport.
On the other hand, when $N < 10$ the honeycomb and square lattices behave differently. 
For AGNR and ZGNR, we have a plateau close to $\gamma=0.5$, which is consistent with L\'evy transport. 
However, for the 2DEG with $N < 10$, the average transmission can no longer be fitted by Eq. (\ref{eq:alpha}). Instead, they follow an exponential, which is compatible with Anderson localization [\onlinecite{sheng2007introduction}].

The results of Fig. \ref{fig:alpharules}(b) confirm the presence of a  L\'evy-to-diffusive transition in graphene nanoribbons imbibed into the EQDC. 
This transition is an exclusive feature of honeycomb lattices, it does not depend on the edge type, and can be tuned by the Fermi energy.
However, in order to access this transition in Fig. \ref{fig:alpharules}(b) we had to produce an ensemble with 6000 samples, which makes the effect experimentally inaccessible. 

Nonetheless, a viable experimental setup can be realized with {only one ELG}, by linearly increasing the electrostatic potential $V$ from $0.23t_0$ to $0.27t_0$, forming a {fictional time series} $T(V)$ with 6000 {time steps} [\onlinecite{pessoa2021multifractal,PhysRevLett.128.236803}].
We calculated the average transmission as a function of length $L$ from the time series and extracted the exponent $\gamma$, shown in 
Fig. \ref{fig:alpharules}(c).
The results of Fig. \ref{fig:alpharules}(b), obtained from 6000 different samples, and those of Fig. \ref{fig:alpharules}(c), obtained from only one sample, are fully compatible, which shows the robustness of the L\'evy-to-diffusive transport transition.

{In order to further  assert the robustness of the transition, we developed a numerical calculation including hopping parameters between second-nearest-neighbors [\onlinecite{RevModPhys.81.109}]. The triangular symbols in Fig. 3(c) correspond to the data for an AGNR including second-nearest-neighbor hopping with energy $t'_0=0.1 t_0$ [\onlinecite{RevModPhys.81.109}]. As expected, we see a L\'evy-to-diffusive transport transition, which shows that the inclusion of  second-nearest-neighbor hopping does not change our results.}
Finally, we also increased the width of the nanoribbons to $W_A= 59.5a_0$ and $W_Z = 60.0a_0$, and implemented the one sample procedure described above to obtain Fig. \ref{fig:alpharules}(d).
{Once again, the L\'evy-to-diffusive transition is visible in both cases, and is not suppressed by an increase in ribbon width, as we asserted above.}

{The experiments in Ref. [\onlinecite{Caridad2016}] demonstrate a high degree of precision and control over the electrostatic cluster deposition process, capable of avoiding overlap between individual clusters. Nonetheless, we expect our results to remain  unchanged for a low cluster overlap density. However, in the case of large cluster overlap, the nanoribbon will approach a situation of constant electrostatic potential in all sites, and we expect its transmission will become similar to the one in Fig. \ref{fig:T_L}(b) (circles).}

\section{Discussion}
A graphene nanoribbon imbibed in an EQDC shows a L\'evy-to-diffusive transport transition as a function of Fermi energy. 
The transport is superdiffusive in the quantum regime, i.e. $N<10$, and diffusive in the semiclassical regime, $N>10$ [\onlinecite{pessoa2021multifractal,Oliveira_2022}]. 
Ref. [\onlinecite{natureEM}] showed  that the transmission fluctuations of graphene as a function of magnetic field are multifractal close to the Dirac point and monofractal far away from it. 
This multifractality has its origin in the quantum correlations induced by an external parameter such as a magnetic field or electrostatic potential, which decreases as the number of channels $N$ (or the Fermi energy) increases [\onlinecite{pessoa2021multifractal}]. 
Nonetheless, graphene presents a linear energy dispersion close to the Dirac point, which preserves chiral symmetry. 
However, both linear dispersion and chiral symmetry are broken by increasing the Fermi energy [\onlinecite{PhysRevE.69.056219,PhysRevB.86.155118,PhysRevB.88.245133}]. 

Given this scenario, what could possibly be the origin of our L\'evy-to-diffusive transport transition?
We list two potential causes: 1) loss of multifractality induced by the increase of the Fermi energy; 2) chiral symmetry breaking also induced by an increase in Fermi energy. 
In what follows, we critically examine both possibilities and conclude that our results are most compatible with the second one.

\begin{figure}
    \centering
    \includegraphics[width=1\linewidth]{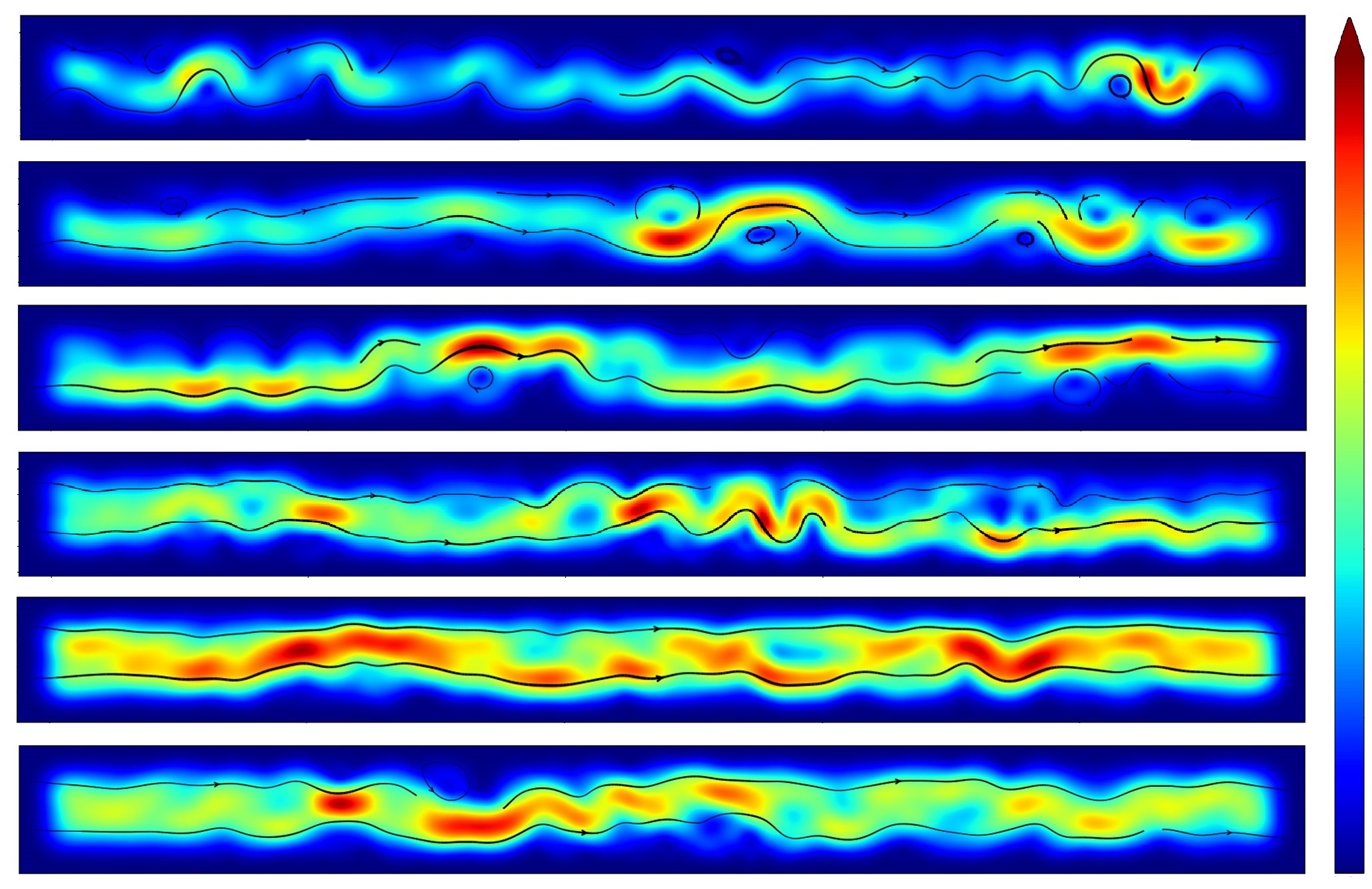}
    \caption{Current density for an AGNR with $L_A=1050.7a_0$ and $N = 1, 2, 5, 10, 20$ and $47$ (top to bottom). Current density increases as the color changes from blue to red.}
    \label{fig:corrent}
\end{figure}

In order to answer the question above, and understand the electronic transport in both regimes, we analyzed the current density of AGNR with $L_A=1050.7a_0$ and $N = 1, 2, 5, 10, 20$ and $47$ (top to bottom), as shown in Fig. \ref{fig:corrent}. 
The current flow deviates from the electrostatic quantum dots in the quantum regime $N<10$, due to the multifractality of the transmission, accompanied by the slow transmission dynamic [\onlinecite{PhysRevB.106.L020201}], i.e., L\'evy transport. 
Besides, the L\'evy transport regime gives rise to vortexes surrounding the smaller quantum dots. 
Those vortexes are an evidence of turbulence-like behavior in an integer quantum Hall transition, as reported in Ref. [\onlinecite{PhysRevLett.128.236803}]. 
Meanwhile, in the semiclassical regime $N>10$, the current flow fills all of the ELG due to the  fast transmission dynamic and destroys the vortexes, compatible with a regular diffusive regime.
We have obtained similar results for a ZGNR.

\begin{figure}
    \centering
    \includegraphics[width=.86\linewidth]{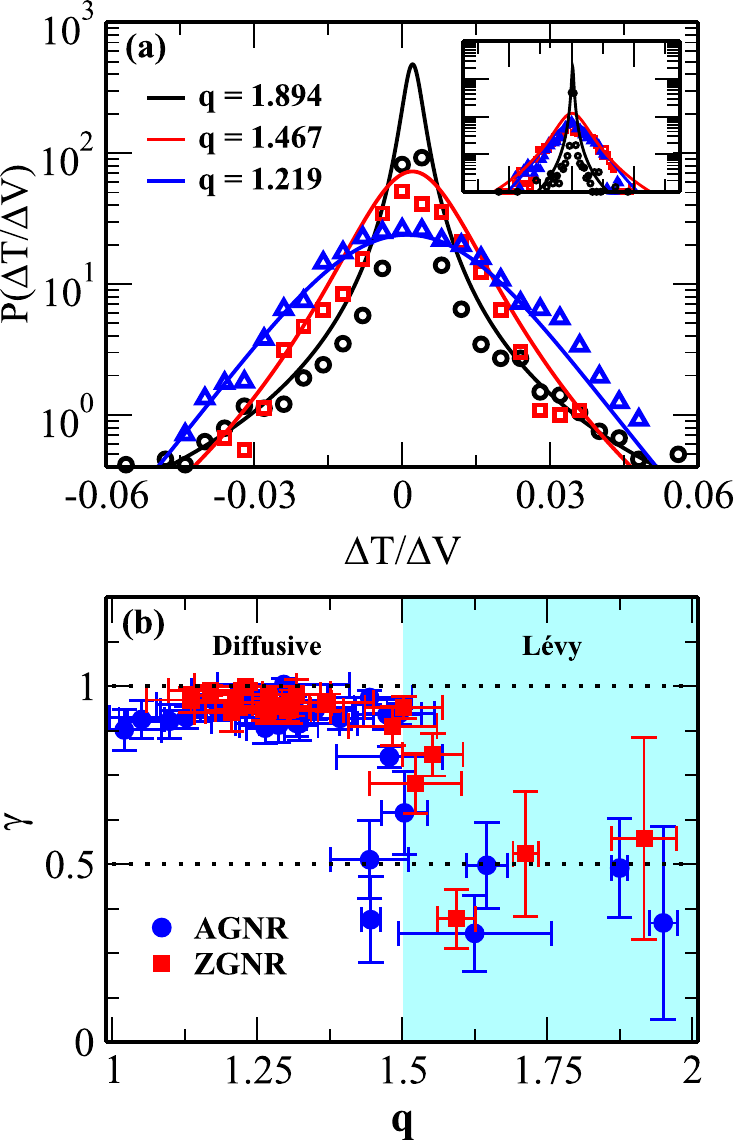}
    \caption{(a) Histogram of the {\it transmission velocity} $\Delta T/ \Delta V$ and its respective q-Gaussian fits from Eq. (\ref{qG}) for AGNR superlattices with $W_A = 49.5a_0$ and $L_A = 1050.7a_0$ with N = 1 (black), 10 (red) and 33 (blue). {The inset shows the histogram of $\Delta T/ \Delta V$ for a 2DEG sample}. (b) The exponent $\gamma$ as a function of the parameter $q$, for AGNR and ZGNR.}
    \label{fig:multifrac}
\end{figure}

We can understand the { slow} and { fast} dynamics, i.e., L\'evy-to-diffusive transition, using the {fictional time series} $T(V)$. 
Fig. \ref{fig:multifrac}(a) shows the histogram of the {transmission velocity}, defined as $\Delta T/ \Delta V$ [\onlinecite{PhysRevLett.79.913}], where $\Delta T = T(V + \Delta V) - T(V)$. 
The data corresponds to an AGNR with $L_A=1050.7a_0$ and $N=1, 10$ and $33$. 
For $N=1$ we have a non-Gaussian distribution (black circles) with a narrow peak around zero and heavy tails, while for $N=10$ and $33$ the distribution approaches a Gaussian.
According to Ref.[\onlinecite{pessoa2021multifractal}], this transition from non-Gaussian to Gaussian behavior is due to the loss of multifractality induced by the increase of the  Fermi energy. 
On the other hand, the inset of Fig. \ref{fig:multifrac}(a) shows that the 2DEG sample also presents a transition from non-Gaussian to Gaussian distribution, which indicates that the loss of multifractality is not an exclusive feature of the honeycomb lattice. 
Hence, we can conclude that the loss of multifractality 
cannot be the cause of our transition, since it is not an exclusive feature of the honeycomb lattice.

Although the loss of multifractality is not the cause of the L\'evy-to-diffusive  transition, we can still use the histograms to characterize it. 
The histograms of Fig. \ref{fig:multifrac}(a) can be fitted by q-Gaussian probability density functions
\begin{equation}
    P(x) = \frac{\sqrt{\beta}}{C_p}[1 + (q-1)\beta(x - x_0)^2]^{\frac{1}{1-q}} \label{qG}
\end{equation}
with
$$C_q = \frac{\sqrt{\pi}\,\Gamma \left (\frac{3-q}{2(q-1)}\right)}{\sqrt{q-1}\,\Gamma\left(\frac{1}{q-1}\right)},$$
where $1 < q < 3$, $\beta$ is a measure of the width of the distribution, and $x_0$ its mean. We remark that Eq. (\ref{qG}) can be
formally derived from a maximization of the Tsallis entropy [\onlinecite{suyari2006mathematical}], and 
note that, when $q \rightarrow 1$, $P(x)$ converges to the Gaussian distribution. 
Therefore, values of $q$ different from 1 can be seen as a measure of non-Gaussianity.

The solid lines in Fig. \ref{fig:multifrac}(a) represent the best fits to Eq. (\ref{qG}), and it is apparent that $q$ decreases as $N$ increases, making it possible to associate the transmission exponent $\gamma$ with the parameter $q$. 
Fig. \ref{fig:multifrac}(b) shows the exponent $\gamma$ as a function of $q$, where the L\'evy-to-diffusive  transition is apparent around $q=3/2$, for both AGNR and ZGNR.
When $q<3/2$ the transmission is diffusive and monofractal, while for $q>3/2$ it is superdiffusive and multifractal. 
{As reported by Refs. [\onlinecite{PhysRevE.64.056134,Pluchino_2007,Gonzalez2017}], the transition from a non-Gaussian to a Gaussian distribution is often associated with a phase transition.
Therefore, Fig. \ref{fig:multifrac}(b) is strong evidence that the L\'evy-to-diffusive transition is a non-equilibrium phase transition.}

The main feature of the graphene honeycomb lattice is its linear energy dispersion close to the Dirac point, while the square lattice has a parabolic dispersion. 
As a consequence of the linear dispersion, charge carriers behave as relativistic massless Dirac fermions and can tunnel through electrostatic potential barriers with null reflection probability, i.e., Klein tunneling effect.
However, the linear dispersion is lost far from the Dirac point, due to an increase of Fermi energy, hence, the charge carriers behave as massive fermions.

We can also interpret this behavior in the context of Random Matrix Theory [\onlinecite{RevModPhys.69.731}].
The electronic transport through the honeycomb lattice in the absence of a magnetic field and spin-orbit interaction is described by the BDI class of Chiral Ensembles, while for the square lattice it is described by the AI class of Wigner-Dyson Ensembles at Cartan's nomenclature [\onlinecite{PhysRevB.86.155118}].
The honeycomb lattice preserves time-reversal, particle-hole, and chiral symmetries at the Dirac point [\onlinecite{PhysRevB.86.155118}]. 
However, far away from it, the BDI class crosses over to the AI class because of the chiral symmetry breaking [\onlinecite{PhysRevE.69.056219,PhysRevB.86.155118,PhysRevB.88.245133}]. 
This explains the different behavior between honeycomb and square lattices in the quantum regime $N<10$, and the similar behavior in the semiclassical regime $N>10$ shown in the Fig. \ref{fig:alpharules}(b) and (c).

Thus, we are led to conclude that the superdiffusive transport is an exclusive feature of ELG in the quantum regime, while diffusive transport is a feature in the semiclassical regime. 
Therefore, the most compelling explanation for the observed L\'evy-to-diffusive  transition in our calculations is the chiral symmetry breaking.
Furthermore, Refs. [\onlinecite{PhysRevLett.111.056801,PhysRevLett.126.206804}] indicate that chiral symmetry breaking has been associated with a phase transition, reinforcing that our transport  transition can be a non-equilibrium phase transition associated with chiral symmetry breaking and, hence, the loss of linear energy dispersion.

Finally, it is interesting to compare our findings with those in Ref.[\onlinecite{bart}]. 
Light has a linear energy dispersion similar to graphene close to the Dirac point. 
Thus, when light is submitted to an optical material in which its transport is governed by L\'evy statistics, it leaves the regular diffusive regime and enters the L\'evy one. 
However, light is not expected to show a L\'evy-to-diffusive  transition because its dispersion is always linear, unlike that of graphene. 

\section{Conclusions}

In conclusion, we proposed an electronic L\'evy glass, analogous to a recent optical realization.
We investigated the transmission of electrons in AGNR and ZGNR in the presence of circular electrostatic clusters, whose diameter follow a power-law distribution. 
We analyzed the effect of the EQDC on the electronic transport of nanoribbons and its diffusion regime, in comparison to a 2DEG.
Our numerical calculations showed that the presence of the clusters induces a transition from L\'evy (superdiffusive) to regular diffusive transport as the Fermi energy increases, which was verified by a q-Gaussian analysis, as shown in Fig. \ref{fig:multifrac}(b).
We conclude that the superdiffusive transport in ELG is an exclusive feature of the low-energy quantum regime, while diffusive transport is a feature of the semiclassical regime.
Therefore, the observed L\'evy-to-diffusive transport transition is mostly likely caused by chiral symmetry breaking once we move away from the graphene Dirac point.
{It would be interesting to investigate how the shape of the EQDC affects the electronic transport behavior and if the superdiffusive regime also appears in the case of non-circular clusters.}

\section*{Acknowledgements}
DBF acknowledges a scholarship from Funda\c{c}\~ao de Amparo a Ci\^encia e Tecnologia de Pernambuco (FACEPE, Grant IBPG-0253-1.04/22).
LFCP acknowledges financial support from Conselho Nacional de Desenvolvimento Cient\'{\i}fico e Tecnol\'ogico (CNPq, Grants 436859/2018 and 313462/2020), and FACEPE (Grant APQ-1117-1.05/22).
ALRB also acknowledges financial support from CNPq (Grant 309457/2021).
We would like to thank C. W. J. Beenakker for bringing Ref. [\onlinecite{PhysRevE.85.021138}] to our attention and for his useful suggestions.






\bibliographystyle{apsrev4-1}
\bibliography{ref.bib}


\end{document}